\documentclass{vldb}
\usepackage{graphicx,epstopdf}
\usepackage{balance}  
\usepackage{enumitem}
\usepackage{hyperref}
\usepackage{caption}
\usepackage{amsmath}
\usepackage[]{algorithm2e}
\usepackage[toc,page]{appendix}
\usepackage{fancyvrb}
\usepackage{xcolor}
\usepackage{graphicx}
\usepackage{url}
\usepackage{balance}  
\usepackage{listings}
\begin{document}


\title{Modelling and Managing SSD Write-amplification}


%
%
%
%

 \numberofauthors{3} 
 \author{
 \alignauthor Niv Dayan\\
        \affaddr{IT University of Copenhagen}\\
        \affaddr{Copenhagen, Denmark}\\
        \email{nday@itu.dk}
 \alignauthor Luc Bouganim\\
        \affaddr{INRIA Paris-Rocquencourt}\\
       \affaddr{Le Chesnay, France}\\
        \email{Luc.Bouganim@inria.fr}
 \alignauthor Philippe Bonnet\\
        \affaddr{IT University of Copenhagen}\\
        \affaddr{Copenhagen, Denmark}\\
        \email{phbo@itu.dk}
 }

\maketitle
\balance

\begin{abstract}

How stable is the performance of your flash-based Solid State Drives (SSDs)? This question is central for database designers and administrators, cloud service providers, and SSD constructors. The answer depends on write-amplification, i.e., garbage collection overhead. More specifically, the answer depends on how write-amplification evolves in time. 

How then can one model and manage write-amplification, especially when application workloads change? This is the focus of this paper. Managing write-amplification boils down to managing the surplus physical space, called over-provisioned space. Modern SSDs essentially separate the physical space into several partitions, based on the update frequency of the pages they contain, and divide the over-provisioned space among the groups so as to minimize write-amplification. We introduce Wolf, a block manager that allocates over-provisioned space to SSD partitions using a near-optimal closed-form expression, based on the sizes and update frequencies of groups of pages. Our evaluation shows that Wolf is robust to workloads change, with an improvement factor of 2 with respect to the state-of-the-art. We also show that Wolf performs comparably and even slightly better than the state of the art with stable workloads (over 20$\%$ improvement with a TPC-C workload). 

\end{abstract}

\section{Introduction} 
\label{sec:introduction}

Solid State Drives (SSDs), based on flash chips, are now the secondary storage
of choice for data intensive applications. Database
systems can now rely on high performance SSDs to store log, indexes and data either on servers or in the cloud.
While SSDs provide increasingly high performance out of the box, 
maintaining high throughput and low latency as the utilization of SSDs increases and despite abrupt changes in the workload remains a challenge. 

On flash chips, data is organized into pages, each of which typically comprises tens of kilobytes of data. Pages reside in erase blocks. 
Before a page is updated, the underlying block must be erased. A software layer called the 
Flash Translation Layer (FTL) manages these constraints by updating pages out-of-place. 
To enable out-of-place updates, SSDs are given {\em over-provisioned} space: a portion 
of the physical storage space which is not exposed to applications, but used by the FTL to accommodate 
the before-images of pages. Eventually,  garbage-collection operations are required to reclaim 
space occupied by before-images. Garbage-collection operations migrate live pages from victim blocks
(to be erased) to other blocks. The flash operations generated by the garbage collector interfere
with application IOs. Such interferences cause 
reduced throughput and variable latency. 
For instance, the NTOSort\footnote{\url{http://sortbenchmark.org/NTOSort2013.pdf}},
with top results in the JouleSort
benchmark, exhibits high variability
for the 1TB sort, when the Samsung 840 Pro SSDs are highly utilized.

The extent of this overhead is captured by {\em write-amplification}, which is the ratio between the number of physical writes that occur inside an SSD to the number of logical writes issued by the application. A way to reduce write-amplification is to increase {\em over-provisioning}\footnote{Other ways to
reduce write-amplification include compression and deduplication. Both of these
methods reduce the volume of data that needs to be stored on SSD, at the cost of
increased CPU utilization. They are orthogonal (and can be combined) with our
proposed contribution.}. Both SSD constructors and application designers tune 
over-provisioning to trade storage capacity for performance
(see for example the note on how over-provisioning is used to maximize lifetime and performance for the Samsung 840 Pro SSDs\footnote{\url{http://www.samsung.com/global/business/semi
conductor/minisite/SSD/global/html/about/whitepaper05.html}}).

There is a direct relationship between write-amplification and over-provisioning. A
greater amount of over-provisioned space reduces the average number of live pages per block, and thus the average
number of pages that need to be migrated when a block is garbage-collected. 

There is also a relationship between write-amplification and the internals of the FTL. 
It has been shown~\cite{stoica} that a way to reduce write-amplification
is to design a block manager, within the FTL, that writes
hot and cold pages on separate flash blocks, thereby essentially partitioning
the SSD into several groups based on the update frequency of the pages they contain. Designing a block manager
that manages data placement across such SSD partitions is still problematic though: 

\begin{itemize}

\item Existing block managers exhibit pathological behaviors when certain types
of changes in the application workloads occur. In the context of a database 
system, examples of such abrupt changes include offline 
index maintenance (which essentially forces database writes 
to swap back and forth between data pages and index pages), 
checkpointing (where writes switch between log and data pages), 
or external algorithms (that force large volumes of writes to
temporary space). Moreover, in cloud
environments SSDs store various databases with different workload
characteristics subject to abrupt changes in the
workload \cite{spikes}.
We show that when a group of pages abruptly changes
temperature, existing FTLs exhibit an elongated period of exceptionally high
write-amplification as over-provisioned space is slow to move among groups.

\item Existing block managers do not adapt well to realistic workloads like TPC-C. 
The reason is that existing methods make assumptions about the relative temperatures 
of pages in different groups, which may be inaccurate. The allocation of over-provisioned 
space based on such assumptions is sub-optimal. 

\item There is no closed-form method for allocating over-provisioned space to different 
groups. Existing methods are either mathematically  complex or based on hill-climbing algorithms, which are computationally
expensive. 

\end{itemize}


In this paper, our contribution is twofold:
\begin{enumerate}
\item We derive a mathematical expression that relates over-provisioning to write-amplification under a uniform workload. 
\item We introduce a new block manager to address the aforementioned problems, called Wolf, or \textbf{Wo}rkload \textbf{L}eveller for \textbf{F}lash. Wolf is able to detect and quickly adapt to changes in workload by pro-actively reallocating over-provisioned space among the groups based on their changing needs. It adapts better to stable workloads by measuring the update frequencies of	groups instead of making assumptions about them. It uses a novel  near-optimal closed-form expression to allocate over-provisioned space to groups.
\end{enumerate}

The rest of the paper is organized as follows. Section 2 explores related work. 
Section 3 introduces a system model for the
analysis, as well as the simulator we later use. Section 4 simplifies an
existing analysis of the relationship between over-provisioning and write-amplification 
with uniform workload, and analyses the behavior of 
garbage-collection efficiency over time in this context. In section 5,  we introduce
Wolf, a new block manager design that adapts better to realistic workloads,
especially ones that dynamically change over time. 
We describe our evaluation of Wolf in Section 6 and conclude in section 7.

\section{Related Work} 
\label{sec:related_work}

We discuss existing work focused on modelling write amplification, and managing
it in the context of a FTL design.

\subsection{Modelling Write-Amplification}

There have been many efforts to relate write-amplification and over-provisioning.
Hu et al.~\cite{write-amp} and Agrawal et al.~\cite{closed} developed
probabilistic models for write-amplification, but they do not fit simulation
results well for all values of over-provisioning. Haas et al.~\cite{limit}
proposed a model for write-amplification that is directly fitted from simulation
results. Bux et al.~\cite{bux} developed two complementary theoretical models
for write-amplification, but they do not give a closed-form expression.

In 2012-2013, several papers~\cite{improved, Desnoyers, stoica} gave an equation
that relates over-provisioning to write-amplification in terms of the Lambert w
function. Stoica et al.~\cite{stoica} even simplified this expression into one
that does not rely on the Lambert w function. The model presented in these works
fits simulation results nicely when the workload is stable and has been running
for a while.

In this paper, we derive a simpler closed-form equation that relates over-provisioning 
to write-amplification. 

\subsection{Managing Write-Amplification} \label{sec:prev_work_managing}

In the design of the Log-Structured File System \cite{log-fs}, the authors
noticed that when writing a stream of logical random writes sequentially in the
physical address space, one can reduce cleaning costs by physically clustering
together logical addresses with the same update frequency. Several works
utilize this idea for flash in order to minimize write-amplification.

Envy \cite{envy} proposed separating the physical space into equally-sized
groups of erase blocks. Pages are migrated to a hotter or colder group based on
their temperatures with the overall goal of equalizing the groups' cleaning
costs, defined as the average cost of cleaning a block multiplied by the
frequency of cleaning operations. DAC \cite{dac} also partitions the physical
space into equally sized groups, but the migration policies are different. A
page is promoted to a hotter group when it is updated and only if the update
distance to the last update is relatively short. On the other hand, a page is
demoted to a colder group during cleaning. ContainerMarking \cite{container} 
partitions the physical space in finer-grained groups, and promotes pages on 
updates and demotes pages based on a statistical model.


 
Recently, Stoica et al.~\cite{stoica} proposed a scheme, that we denote Frequency-based
Data Placement (FDP) in the rest of the paper, whereby the number and
sizes of groups adapt to the workload. In FDP, cleaning is performed independently in
each group using a least-recently-cleaned (LRU) policy. Several methods are
proposed for allocating over-provisioned space to the different groups in order 
to minimize overall write-amplification, yet no closed-form is given. FDP outperforms 
existing FTL designs theoretically and under realistic workloads.

In this work, we propose Wolf, which improves on FDP~\cite{stoica} in several ways. It
adapts better to realistic workloads since the update frequencies of groups are
continuously measured and adapted. It avoids pathological behabiors that FDP is subject to when application workloads change. It uses a greedy cleaning policy in groups, as we identified
scenarios in which the LRU policy is suboptimal. Finally, it uses a simple
closed-from expression to decide how to allocate over-provisioned space to the
different groups.



\section{System Model} 
\label{sec:system_model}

We rely on the following SSD model throughout this paper.
There are $B$ pages in a block, and $K$ blocks
in the SSD. The SSD is partitioned into a number of logical units called LUNs,
on which flash operations can occur in parallel. Several LUNs are connected to
the SSD controller through channels. Communication between the controller and
the LUNs through the channel is serial.

The size of the logical address space in pages is $LBA$, and the number of physical
pages is $PBA$. The amount of pages for over-provisioned 
space is $OP = PBA - LBA$. Over-provisioning is captured by the ratio $LBA/PBA$.  The notations used for the analysis 
are summarized in Table \ref{tab:notations}. 

\begin{table}[t]
\begin{center} 
    \begin{tabular}{ | l | p{6cm} |} 
    \hline
    \textbf{Notation} & \textbf{Description} \\ \hline
    $B$ & Number of pages in a flash block  \\ \hline
    $\delta$ & The fraction of pages in a flash block that are on average migrated in one garbage collection operation \\ \hline
    $LBA$ & Logical address space size in flash pages \\ \hline
    $PBA$ & Physical address space size in flash pages \\ \hline
    $OP$ & The number of over-provisioned flash pages. Note that $OP = PBA - LBA $ \\ \hline
    $K$ & The number of blocks in the SSD. Note that $K = \frac{PBA}{B}$ \\ \hline
    \end{tabular}
\end{center}
\caption{System Model}   
\label{tab:notations}
\end{table}

We assume the FTL uses a page-level mapping scheme to handle out-of-place
updates. This means there are LBA mapping entries, and when a page is updated,
the physical address corresponding to the page's logical address is changed.
The mapping table is either stored in the host's RAM or on flash
memory \cite{dftl}. 

In our baseline block manager, a page update is written to some non-busy LUN
that has a block with free pages. As the LUN runs out of blocks with free
space, cleaning operations are triggered on it to clear space for new writes.
Garbage-collection works independently for each LUN. It selects a victim,
migrates any live pages to other blocks (which may be on other LUNs), and
finally erases the block. The block is then returned to the pool of free
blocks for that LUN.

The garbage-collection policy within a LUN can be {\em LRU} or {\em greedy}.  The LRU
policy targets the block that was erased the longest time ago. The greedy
policy keeps track of the number of live pages in each block and targets the
block with the least number of live pages. 

We use the SSD simulator EagleTree \cite{EagleTree} to validate analytical
results and evaluate system designs. EagleTree is an open-source and
extensible simulation framework for the entire IO stack running in virtual
time. We used the above-mentioned policies and the default parameters in table
\ref{tab:sim-parameters} for all simulations unless otherwise mentioned\footnote{The default
settings result in an SSD of 16GB. We show throughout the paper that increasing 
SSD capacity (i.e., the number of channels, the number of LUN per channels, the 
number of blocks per LUN or page size) either has no impact on our analysis, or
that it actually amplifies the benefits of our approach. This setting is thus conservative
and it allows us to conduct a thorough exploration of the design space in simulation.}. We
implemented Wolf on top of the existing simulator, 
available on Github\footnote{\url{https://github.com/ClydeProjects/EagleTree}}.

\begin{table}[t]
\begin{center}  
    \begin{tabular}{ | l | p{2,5cm} |} 
    \hline
    \textbf{Parameter} & \textbf{Default Value} \\ \hline
    Channels & 4  \\ \hline
    LUNs per channel & 2  \\ \hline
    Blocks per LUN & 1024 \\ \hline
    Pages per block (B) & 128 \\ \hline
        Page size & 16 KB \\ \hline
    LBA / PBA & $70\%$ \\ \hline
    \end{tabular}
\end{center}
\caption{Default simulation parameters}
\label{tab:sim-parameters} 
\end{table}


\section{Write-Amplification} 
\label{sec:write_amplification_for_uniform_workloads}

We now examine the relationship between over provisioning and write-amplification. 
We focus on a uniformly distributed random workload, where an application write targets any logical address with an equal probability. 



\subsection{The lifetime of a block}\label{sec:block-lifetime}

As a first step, we model the lifetime of a block. Suppose we have just finished writing a block. 
How many pages G do we expect to remain valid after X application 
writes\footnote{We assume that LBA is much bigger than B, which means the likelihood 
that a page in the block was invalidated before we finished writing the block is negligible. 
Thus, we assume all pages in this block are initially valid.}? 

Since the workload is uniform, the probability that any page update targets some page in the block is $\frac{B}{LBA}$. Thus, the number of page updates before some page is invalidated in B is geometrically distributed with mean $\frac{LBA}{B}$. In other words, $\frac{LBA}{B}$ page updates occur on average before the first page in the block is invalidated. After this event, there are $B-1$ live pages in the block. Analogously, the expected number of page updates before the next page in the block is invalidated is $\frac{LBA}{B-1}$. The third page is invalidated after an expected number of $\frac{LBA}{B-2}$ writes, and so on. The expected number of page updates until there are no live pages in the SSD can be simplified using Euler's approximation for the sum of the harmonic series up to n elements, where $\gamma$ is the Euler-Mascheroni constant.

$$
\frac{LBA}{B} + \frac{LBA}{B-1} + ... + \frac{LBA}{1} = LBA \sum_{i=1}^{B}\frac{1}{i} = LBA ( \ln(B) + \gamma )
$$

More generally, suppose that after X updates there are on average $G$ live pages left in the block. It is easy to see that:
 
\begin{equation} \label{eq1}
\begin{split}
X & = LBA \sum_{i=1}^{B}\frac{1}{i} - LBA \sum_{i=1}^{G}\frac{1}{i} \\
& = LBA ( \ln(B) + \gamma ) - LBA ( \ln(G) + \gamma ) \\
& = LBA \ln(B/G)
\end{split}
\end{equation} 

Expressed in terms of G, we get: 

\begin{equation} \label{eq:how_long_til_equib}
\begin{split}
G = B \cdot e^{-X/LBA}
\end{split}
\end{equation} 

\subsection{Equilibrium}\label{sec:unif_equib}

We now study the relationship between over-provisioning ($LBA/PBA$) and the average fraction of migrations per cleaning operation ($\delta$). 


We start by asking a simple question: what is the expected number of cleaning operations throughout the SSD between two times that the same block is cleaned? Under the LRU policy, the number is K by definition. Under the greedy policy, the number is K on average. 


Let us now ask how many application page updates are expected between two times the same block is cleaned. By the assumption of being in equilibrium, each cleaning operation involves $B \cdot \delta$ migrations. Conversely, each cleaning operation clears space for $B \cdot (1 - \delta)$ new application writes. Since K blocks are cleaned on average between two times that the same block is cleaned, and since each of them accommodates for $B \cdot (1 - \delta)$ application writes, then $K \cdot B \cdot (1-\delta)$ page updates occur on average 
before two times that the same block is garbage-collected. We substitute this expression for X in equation \ref{eq1}. 

$$
LBA \cdot \ln(\frac{B}{B \cdot \delta}) = K \cdot B \cdot (1-\delta) \\
$$

Plugging in $\frac{PBA}{B}$ for K, simplifying and rearranging, we get: 

\begin{equation} \label{eq:eq_6}
\frac{LBA}{PBA} = \frac{\delta - 1}{\ln(\delta)}
\end{equation} 

We can relate $\delta$ to write-amplification using the equation $WA = 1/(1 - \delta)$. This is a characterization of write-amplification at equilibrium.

Figure \ref{fig:verif} plots equation \ref{eq:eq_6}. In appendix \ref{App:AppendixA}, we show how to express equation \ref{eq:eq_6} in terms of $\delta$. By so doing, we also show that this equation is in fact equivalent to the findings of previous analyses provided in \cite{closed,stoica,Desnoyers}. The interesting part of our analysis is that the derivation and form of the equation are simpler. 


\begin{figure}
\includegraphics[scale=0.25]{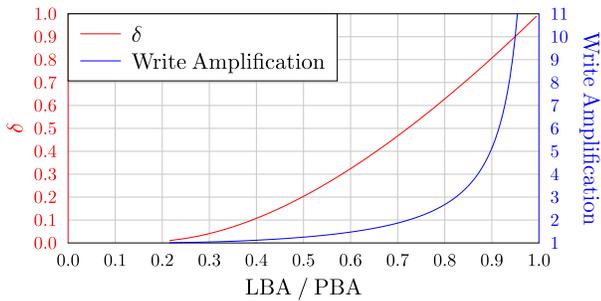}
\captionsetup{margin=10pt,font=small,labelfont=bf}
\caption{ How $LBA / PBA$ affects $\delta$ (left axis) and write amplification (right axis) in equilibrium, assuming a uniform workload. }
\label{fig:verif}
\end{figure}

\section{Wolf: Workload Leveller for Flash} \label{sec:wolf}

We have seen how write-amplification is linked to over-provisioning for random writes uniformly distributed across the whole SSD.  Now consider that the logical address space is partitioned into $n$ groups denoted as $g_1, g_2, ... , g_n$, that each group has a different size $s_1, s_2, ... , s_n$ measured in flash pages, and that each group may have a different update frequency $p_x$ defined as the probability that an incoming application update targets a page in group x. 

Previous works \cite{envy, dac, container, stoica} have shown that storing pages with different update frequencies in different groups of physical blocks leads to reduced write-amplification in this context. Existing approaches thus conceptually partition the SSD into n virtual SSDs. 

The problem is then to allocate the over-provisioned space across virtual SSDs in a way that minimizes write amplification. Existing solutions do not always adapt well to realistic workloads, and they exhibit poor performance under certain types of workload changes that are typical of database workloads. In this Section, we introduce Wolf, a block manager designed to deal with these issues. 

But first, let us describe the problems of existing approaches in more detail. We base our study on the block manager from Stoica et al.~\cite{stoica}, that we denote {\em FDP}, which has been shown to dominate the state-of-the-art. The groups in this scheme have a fixed order, and it is assumed that the $i$th group contains hotter pages than the $(i-1)$th group. A page is promoted/demoted to a hotter/colder group on an update/migration if it is deemed too hot/cold relative to the other pages in its group.

Problems arise if the update frequency of a group abruptly changes. The essential problem is that over provisioned space is slow to move among groups, which leads to a potentially long period of suboptimal write-amplification. 

For example, Figure~\ref{fig:mass_migrations_prob} (in Section 6) illustrates a scenario using FDP where we initially have two groups $g_{cold}$ and $g_{hot}$ where $s_{cold} = s_{hot}$, $p_{cold} = 10\%$ and $p_{hot} = 90\%$. At some point, the update frequencies of the groups are abruptly swapped. Pages in $g_{cold}$ are now hot and flow into $g_{hot}$ very quickly as they are updated. On the other hand, pages originally in $g_{hot}$ are now cold, and they reside in blocks that occupy a lot of over-provisioned space. Ideally, these pages should be quickly demoted and compacted into $g_{cold}$ in order to free the underlying over-provisioned space, so that it can now benefit the pages that turned hot. However, there is no mechanism to do this quickly. The over-provisioned space is only freed as the original blocks in $g_{hot}$ are cleaned, and the amount of time this takes depends on the garbage-collection policy. Indeed, in figure~\ref{fig:mass_migrations_prob}, write-amplification after the temperature swap is slow to converge again. 

The question we ask is thus the following: how can we facilitate a speedy reallocation of over-provisioned space among groups in order to minimize overall write-amplification as fast as possible? 

The scheme we propose, Wolf, addresses these concerns. It is able to quickly detect changes in workload and quickly move over-provisioned space among groups in response. A key design principle is that when a change in workload occurs, physical space rather than logical pages (as in existing schemes), should move among groups. Wolf is different from existing schemes in the following ways. 

It continuously records run-time statistics about the sizes and update frequencies of groups. These statistics serve three purposes. (1) They allow adjusting the relative ordering of groups as the workload changes. (2) They allow determining when it is worthwhile creating or merging groups as the workload changes. (3) They serve a reliable input to determine how to reallocate the over-provisioned space among groups as the workload changes. 

Wolf uses a novel near-optimal closed-form method to compute how to allocate over-provisioned space among the groups. Compared to existing methods, the closed-form method constitutes a good compromise between cheapness, accuracy and simplicity. 

Wolf continually identifies groups with excess over-provisioned space relative to the optimum, compacting pages in those groups on fewer physical blocks, and donating the redeemed blocks to groups with a deficit of over-provisioned space. This is achieved through \textit{movement operations}. Within a group, Wolf uses the greedy garbage-collection policy since our analysis showed that it significantly outperforms the LRU policy under certain workload changes.

We first describe the key data structures and skeleton subroutines of Wolf. We then describe its novel aspects in more detail.


\subsection{Algorithm Outline and Key Data Structures}

Wolf's skeleton is structured around three key subroutines, and the notion of intervals of application writes. The first subroutine handles write operations and updates statistics about each group within an interval. The second subroutine handles the completion of an interval. In this subroutine, updated statistics about groups' sizes and update frequencies are used to (1) re-sort the relative order of group, (2) consider whether or not to create or merge groups, (3) recompute the optimal allocation of over-provisioned space among groups, and (4) consider moving over-provisioned space among groups. The third subroutine handles an erase operation. It assigns a redeemed block to a group with an over-provisioned space deficit, and considers triggering more over-provisioning reallocation operations. We now explain these subroutines in more detail. 


Listing \ref{lst:writes} shows how writes are handled. When a write arrives, we determine which group it currently belongs to by (1) looking up the corresponding physical address in the usual logical to physical mapping table (orthogonal to Wolf), and then (2) consulting a data structure called the \textit{Blocks to Groups Map (BGM)}, which keeps track of which physical blocks belong to which group. Now that we know the page's current group, we consult a temperature detector module (TD) to determine if the page should remain in its current group or be demoted/promoted to a colder/hotter group. When we have a target group on which to write the page, we find a free page in the group and execute the write. Finally, we update statistics about the sizes of groups and the number of writes targeting the groups per interval. The routine concludes by checking if the target group is nearly out of free pages, in which case garbage-collection is invoked within it. Non-trivial parts of all subroutines are marked in red and described in more details in later sections. 

\lstset{morekeywords={MT,BGM, TD, flash, GROUPS, SGV}}
\lstset{emph={write},emphstyle=\textit}
\lstset{emph={find_target, garbage_collect},emphstyle=\textcolor{red}}

\begin{lstlisting}[frame=single, numbers=left,caption={ handle write},breaklines=true,label={lst:writes},columns=fullflexible]
  input: write
  logi_addr = write.logi_addr
  phys_addr = MT[logi_addr]
  block_id = phys_addr.block_id
  group_id = BGM[block_id]
  curr_group = GROUPS[group_id]
  new_group_id = TD.find_target(group_id, logi_addr) 
  new_group = GROUPS[new_group_id]
  new_phys_addr = new_group.find_free_addr()
  write.new_phys_addr = new_phys_addr
  flash.schedule(write)
  curr_group.size--
  new_group.size++
  if (write is not a garbage-collection operation)
      new_group.interval_writes++
  if (new_group.num_free_pages < B)
      new_group.garbage_collect()
\end{lstlisting}

In principle, the length of an interval, denoted as $h$, captures a trade-off between responsiveness to changes in workload and the computational expense of subroutine \ref{lst:interval}. Fortunately, subroutine \ref{lst:interval} is not expensive since the component that would usually be expensive in it, namely the method for recomputing the allocation of over-provisioned space, is non-iterative and therefore extremely cheap. Thus, we can set $h$ to be small (in our implementation $h = LBA \cdot 0.001$\footnote{The parameters we define in this Section remained fixed throughout the experiments presented in Section 6. In this paper, we do not present a thorough exploration of the performance impact of these tuning parameters. This is a topic for future work.}) 

Let us now describe subroutine \ref{lst:interval}. In lines 2 and 3, each group's update frequency is updated based on the proportion of writes that targeted that group within the interval that finished. Note that $a$ is a constant that controls the weighting of a group's long-run measured update frequency versus the update frequency in the interval that just finished (we set $a$ to $h \cdot 3$). In line 4, the allocation of over-provisioned space for a group is recomputed. 

In line 7, we sort a data structure called the \textit{Sorted Groups Vector (SGV)} based on updated statistics. This data structure maintains the relative order of groups based on the notion of \textbf{hit rate}, which we define for group $g_x$ as $p_x / s_x$, its update frequency over its size. As we will later see, this structure is used by the temperature detector, and by the policy that decides whether to create or merge groups. In lines 8 and 9, we trigger routines that consider creating or merging existing groups, and whether or not to start reallocating over-provisioned space among group.

\lstset{emph={allocate_over_prov, merge_or_create_groups, consider_movement_operations},emphstyle=\textcolor{red}}

\begin{lstlisting}[frame=single, numbers=left,caption={ handle interval completion},breaklines=true,label={lst:interval},columns=fullflexible]
  for (group x in GROUPS) {
     U = group.interval_writes / h
     p_x = p_x * (1 - a) + a * U
     group.over_prov = allocate_over_prov(p_x, s_x)
     group.interval_writes = 0
  }
  SGV = sort_by_hit_rate(GROUPS) 
  merge_or_create_groups()
  consider_movement_operations()
\end{lstlisting}

The last subroutine of Wolf handles erase operations. It identifies the group with the greatest over-provisioning deficiency and assigns the erased block to it. After this, we consider reallocating more over-provisioned space by triggering additional movement operations. 

\begin{lstlisting}[frame=single, numbers=left,caption={ erase},breaklines=true,label={lst:erase},columns=fullflexible]  
  flash.schedule(erase)  
  find group x that maximizes (s_x, OP_x)
  block_id = erase.block_id
  BGM[block_id] = group.id
  consider_movement_operations()
\end{lstlisting}

With Wolf's skeleton in mind, let us reiterate why it adapts better to changes in workload. Firstly, Wolf detects changes in workload through statistics. It is highly responsive to changes in workload since intervals are short. As groups change in update frequency relative to each other, their order is re-sorted, and the allocation of over-provisioned space among them is recomputed and pro-actively adjusted. This is in sharp contrast to existing schemes, whereby the ordering of groups is fixed, and write-amplification is controlled by moving pages among the groups, as opposed to by moving over-provisioned space among them.

An additional advantage of Wolf over existing schemes is that its statistics provide a more accurate input to the over-provisioning allocation strategy. FDP makes assumptions about the update frequencies of the groups based on their relative order, but does not measure them. Thus, Wolf is able to adjust better to a stable workload.

The next sections explain non-trivial parts of Wolf, marked in red in the above subroutines. In section \ref{sec:Dynamic_Group_Management}, we describe the policy for creating and merging groups. In section \ref{sec:Over-provisioning_Redistribution}, we describe the policy for reallocating over-provisioned space among groups using \textit{movement operations}. In section \ref{sec:cleaning_within_groups}, we discuss the garbage-collection policy within groups, and in section \ref{sec:overprob_alloc} we describe the novel closed-form method for a near-optimal allocation of over-provisioned space among groups. Finally, in sections \ref{sec:temp_det}, \ref{sec:Garbage-Collection-and-Parallelism} and \ref{sec:wear-leveling}, we discuss compatibility with orthogonal concerns of temperature detection, parallelism and wear-levelling respectively.

\subsection{Groups Creation And Merging Policies} \label{sec:Dynamic_Group_Management}

Let us describe the policy for creating or merging groups invoked in subroutine \ref{lst:interval}. Wolf is initialized with a minimum of two groups. A page that is written for the first time is always assigned to the coldest group. A hotter additional group is created when the following conditions hold: (1) the group with the highest hit rate in SGV must have at least F pages, and (2) the ratio between the hit-rates of the two hottest groups in SGV must be at least Q. Rule 1 is meant to avoid the creation of an excessive number of very small groups, and rule 2 is meant to ensure that the creation of a new group is motivated by a real temperature skew in the workload. 

We set the constant F in rule (1) to be the number of LUNs in the SSD multiplied by the number of pages in a block. Thus, each group has at least one block in each LUN. This ensures natural load balancing across the parallel architecture of the SSD.

An implication of rule (2) is that the hit rate of groups increases exponentially for a stable workload. Indeed, \cite{stoica} showed analytically that the update frequency ratio among pages in the same group may vary by up to 2 without incurring any significant penalty. Thus, we set the constant Q in rule (2) to two. Note that the exponential increase in the hit rates of groups means we can handle very skewed workloads with relatively few groups. 

When a new hot group is created, it takes time for pages to flow into it and for its long-term update frequency $p_x$ to adjust and stabilize. We give it time by fixing its position in the SGV as well as banning the creation or merging of any additional hot group for the next w intervals ($w = 50$ in our implementation). 

If the update frequency of two adjacent groups in SGV diverges and the hotter one is hotter by a factor of more than $Q \cdot 2$, we create a new empty group in the middle, again fixing its location for w intervals to allow it to stabilize. 

If two adjacent groups in SGV converge in terms of hit rate for over $w$ intervals, they are merged. Moreover, if the number of pages in a group drops below F, we merge the group with one of its adjacent groups in SGV. Note that a merge is logical, as it only involves consolidating the metadata about the groups.  

The implication of our groups' creation and merging policy is that when a workload is highly skewed, the number of groups would tend to oscillate. This is indeed what we observe with a TPC-C workload in section \ref{sec:realistic}. 

This policy improves on FDP in the sense that FDP does not have a policy for merging groups, or to control that the creation of a new group is motivated by a genuine skew in the workload as opposed to randomness. Thus, the number of groups in FDP tend to grow over time. 



\subsection{Movement Operations}\label{sec:Over-provisioning_Redistribution}

In both subroutines \ref{lst:interval} and \ref{lst:erase}, a procedure is called to consider reallocating physical space among the groups. It work by scanning the metadata for each groups. It identifies groups that have a block-surplus relative to the dictated amount of over-provisioned space the group should have. It then triggers garbage-collection operations within such groups. The redeemed blocks are donated to groups with a block-deficiency relative to the dictated allocation, as shown in subroutine \ref{lst:erase}.



There is an interesting cost-benefit question regarding how greedily to trigger movement operations. On the one hand, movement operations constitute a clear cost in terms of migrations. On the other hand, movement operations are an investment in future performance, as they help reducing the overall write-amplification. Thus, an interesting question is how rapidly to issue movement operations. We investigated different strategies of pacing movement operations, but found that issuing movement operations as greedily as possible with no restrictions always minimized the overall number of migrations. This indicates that the investment in movement operations is always worthwhile.

In contrast to Wolf, in FDP a group may only be donated blocks from the two groups adjacent to it, namely the next colder and next hotter groups. This restricts the adaptivity of the scheme to changes in workload, as a block may need to cross multiple groups to reach its target. In contrast, Wolf has the flexibility of being able to move blocks between any groups.





\subsection{Cleaning Policy within a Group}\label{sec:cleaning_within_groups}

A garbage-collection operation is triggered in a group when it has less than B free pages. The question we now address is how to choose a victim block. We compare two policies: LRU and greedy.

Previous work~\cite{stoica} showed that the difference between the greedy and LRU policies in equilibrium is not great (though it does increase slightly for higher values of $LBA / PBA$). As LRU is simpler to implement, state-of-the-art FTLs such as FDP~\cite{stoica} use the LRU policy. However, the heuristic of the LRU policy, namely that the block that was cleaned the longest time ago contains the fewest live pages, may sometimes underperform, particularly due to movement operations, which were introduced in the last section. 

Movement operations pack the pages of groups with block-excess more compactly on fewer blocks. Suppose that many of them occur rapidly in a group relative to the size and update frequency of the group. By the time they are finished, a significant number of blocks in the group will be completely full. In such a scenario, the block that has been least recently erased holds statistically the same number of live pages as the other blocks in the group, so the LRU heuristic fails. 

\begin{figure} 
\includegraphics[scale=0.25]{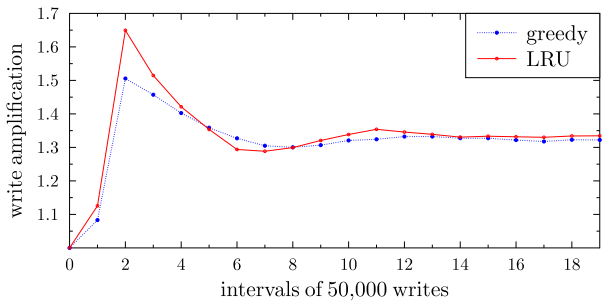}
\captionsetup{margin=10pt,font=small,labelfont=bf}
\caption{The greedy vs. LRU policies}
\label{fig:greedy.vs.lru}
\end{figure}

To demonstrate this, we made a micro-experiment using Wolf whereby the workload is divided into two groups $g_1$ and $g_2$ where $s_1 = s_2$, $p_1 = 100\%$ and $p_2 = 0\%$. Initially, $g_1$ has most of the over-provisioned space and $g_2$ has almost none. We then abruptly swap the groups' update frequencies. Movement operations are issued, rapidly compacting the pages of group 1 into few physical blocks and moving the redeemed blocks to group 2. This creates a vast number of filled blocks with the same number of live pages in group 1. We then swap the update frequencies of the groups again. Cleaning operations are triggered in group 1, but since the blocks in this groups have the same number of live pages, the LRU policy is unable to choose a block with exceptionally few live pages. Write-amplification after the second swap is shown in Figure~\ref{fig:greedy.vs.lru}. Approximately $15\%$ more migrations take place in the LRU policy compared to the greedy policy until write-amplification converges. 
  
Traditionally, the argument used against the greedy policy has been that its computational cost is high, particularly the process of choosing a victim. However, victim selection does not need to consider all blocks in the SSD. It is enough to consider only the blocks in a particular group within a particular LUN. As there are usually tens of LUNs in an SSD, this narrowing of the search allows reducing the cost of victim-selection by an order of magnitude. 

\subsection{Over-Provisioning Allocation} \label{sec:overprob_alloc}

We now describe the closed-form method for determining how to allocate over-provisioned space among groups. Let $OP_x$ be the over-provisioned space given to group x. The total amount of physical space consumed by group x is $(s_x + OP_x)/B$ physical blocks. Let $\delta_x$ be the fraction of pages that need to be migrated per cleaning operation in group x. Using equation \ref{eq:eq_6}, we can relate $s_x$, $OP_x$ and $\delta_x$ for any group x:

\begin{equation} \label{eq:eq_group}
\frac{s_x}{s_x - OP_x} = \frac{\delta_x - 1}{\ln(\delta_x)}
\end{equation} 

Let us denote $WA(s_x, OP_x)$ as the write-amplification of group x, which is a function of the size and over-provisioned space allocated to that group. Overall write-amplification is the sum of the write-amplifications of all groups weighted by their update frequencies. 

\begin{equation} \label{eq:total_write_amp}
\begin{split} 
\text{WA} & = \sum^n_{i=1} p_i \cdot WA(s_i, OP_i) 
\end{split}
\end{equation}

Allocating the over-provisioned space among the groups so as to minimize write amplification is an optimization problem. Although~\cite{stoica, Desnoyers} proposed methods for finding the optimal allocation, no closed-form method was proposed. Moreover, methods explored in these works are either expensive (e.g. relying on iterative optimization algorithms), of questionable accuracy (polynomials of best fit that hold the relative update frequencies of groups as a constant), or of questionable adaptability to workload changes (i.e. they only support movement of blocks between adjacent groups). It is important that the over-provisioned allocation strategy should be both accurate and cheap to compute because it should be invoked frequently, especially with a changing workload. We now present a novel closed-form approach. 

Our key insight is that the optimal allocation of over-provisioned space exists on a plateau defined by a few key points in the optimization space. We find two such points by considering two hypothetical suboptimal policies and then merging their solutions. The first policy assigns over-provisioned space to groups only based on the group's size, and the other assigns over-provisioned space only based on the group's update frequency. We can then approximate the location of the optimum based on these two key points. 

\subsubsection{Considering just Size}

The first policy performs garbage-collection greedily across groups. It always picks as a victim the block with the least number of live pages in the SSD, regardless of which group it belongs to. A redeemed block is allocated randomly to any group that has less than B free pages. 

Let us reason about equilibrium in this kind of system. Suppose there are two groups $g_1$ and $g_2$ with different cleaning efficiencies such that $\delta_1 < \delta_2$. The cleaning policy will only target blocks from group $g_1$ since that's where the cheapest blocks to claim are. However, writes are still taking place in group $g_2$. As $g_2$ runs out of space, some of the blocks garbage-collected from group $g_1$ will flow to group $g_2$. This will increase $\delta_1$ and decrease $\delta_2$, until they equalize. At this point, the blocks from both groups will be just as expensive to garbage-collect. 

This equalization principle holds regardless of the number of groups, their sizes or their update frequencies. Eventually, cleaning efficiency among all groups equalize such that: $\delta_1 = \delta_2 = ... = \delta_n$. This also means that the amount of over-provisioned space in each group reaches a fixed point (it may oscillate, but the cleaning policy will cause it to converge again). So, even though this is an open system, whereby blocks can flow among groups, in equilibrium the amount of over-provisioned space in each group can be assumed to be fixed, and so each group can be analysed as a closed system. 

By equation \ref{eq:eq_group}, the fact that cleaning efficiency for all groups in equilibrium is equal implies that the following holds:

$$
\frac{s_1}{s_1 - OP_1} = \frac{s_2}{s_2 - OP_2} = ... = \frac{s_n}{s_n - OP_n}
$$

It is therefore easy to see that for any group $g_x$: 

$$
\frac{s_x}{s_x - OP_x} = \frac{s_1 + ... + s_n}{s_1 + ... + s_n + OP_1 + ... + OP_n} = \frac{LBA}{PBA}
$$

With some rearranging, the the exact amount of over-provisioned space allocated to group x in equilibrium is the following expression, which only depends on its size. We simplify by letting $V = \left( \frac{PBA}{LBA} - 1 \right)$.

\begin{equation} \label{eq:OP_space_per_group}
OP_x = s_x \cdot \left( \frac{PBA}{LBA} - 1 \right)  = s_x \cdot  V 
\end{equation} 

\subsubsection{Considering just Update Frequency}

The second hypothetical policy we consider assigns over-provisioned space directly based on each group's update frequency. It is solely based on the observation that a group with a higher update frequency should have more over-provisioned space. Thus, the overall over-provisioned space is partitioned to groups based on each group's update frequency, and completely disregarding their sizes. 

\begin{equation} \label{eq:OP_space_per_group_prob}
\begin{split}
OP_x & = p_x \cdot OP \\
& = p_x \cdot \left( V^{-1} + 1 \right) \\
\end{split}
\end{equation} 

Note that this policy directly fixes the amount of over-provisioned space in each group, whereas in the previous one the amount of over-provisioned space per group was determined indirectly through a convergence process. 

\subsubsection{Mixing the expressions} \label{sec:mixing}

Let us now combine the above two policies. Suppose we wish to find a near-optimal allocation of over-provisioned space for the groups. We get two different values from equations \ref{eq:OP_space_per_group} and \ref{eq:OP_space_per_group_prob}. The actual optimum must lay somewhere in-between those values. In fact, it turns out that their average is extremely close to the optimum:

\begin{equation} \label{eq:OP_mixed}
OP_x = \frac{s_x \cdot V + p_x \cdot \left( V^{-1} + 1 \right)}{2}
\end{equation} 

\begin{figure} 
\includegraphics[scale=0.7]{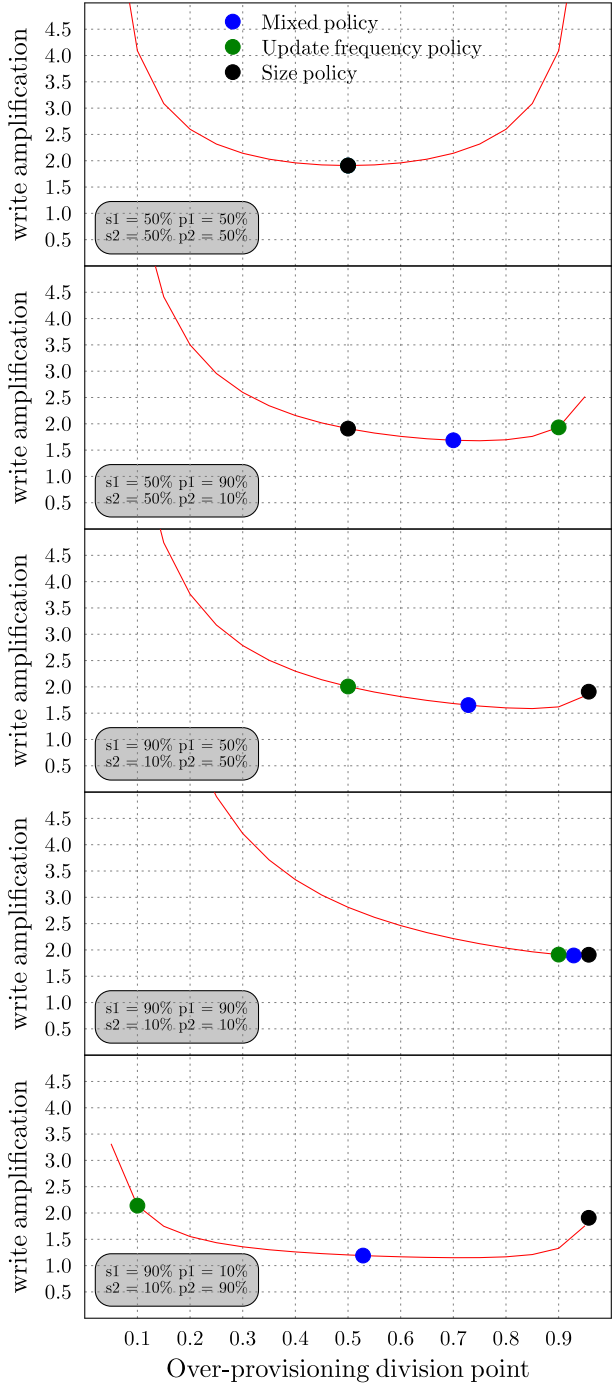}
\captionsetup{margin=10pt,font=small,labelfont=bf}
\caption{ The write-amplification optimization space for several different 2-modal workloads. The fraction from 0 to the blue dot is the amount of over-provisioned space given to group 1, and the rest is given to group 2. The green and black dots corresponds to the point of division based on update frequency or group size alone. The blue dot is their average. }
\label{fig:2_modal}
\end{figure}

This is shown for several 2-modal workloads in Figure~\ref{fig:2_modal}. The red line shows how the division-point of the over-provisioned space among the groups affects overall write-amplification. The dots along the line give the division-point given by the 3 different policies. 

We evaluated formula \ref{eq:OP_mixed} through a brute-force exploration of the workload space. We partition the logical address space and update frequency space into Q equally sized chunks of sizes $LBA/Q$ and $1/Q$ respectively. Each group gets at least one chunk of the logical address space and update frequency space. Under these constraints, we test all possible unique workload configurations. We vary the number of groups from 2 to 9. We repeated all experiments for 4 different over-provisioning values, 0.6, 0.7, 0.8 and 0.9. Note that Q controls how skewed the workload can be. The bigger it is, the more extreme the differences between groups in terms of update frequencies and sizes can be. We used two different values of Q, 10 and 20. All in all, our exploration captured a little over a million different workload configurations. 

Note that the space of workloads covered in our exploration covers a TPC-C workload. Figure \ref{fig:skew} shows an update histogram of the core data of a TPC-C workload. We see that pages are divided into two clusters in terms of temperature. The hotter cluster's peak is approximately 8 times hotter than the other cluster, and both clusters are similar in size. It is easy to see that the above brute-force approach covers such a workload configuration. We demonstrate this empirically in the evaluation. 

The baseline we used to compare the closed-form method is against is the hill-climbing algorithm similar to the one given in~\cite{stoica}, which always finds the optimum because the optimization space for write-amplification is convex. 

\begin{figure}[t]
\includegraphics[scale=0.25]{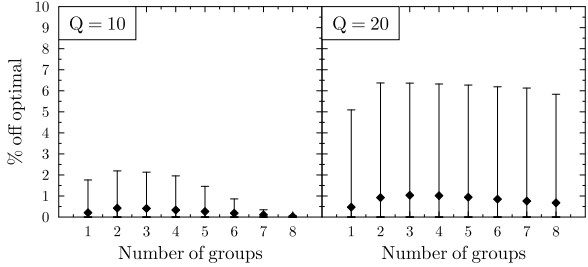}
\captionsetup{margin=10pt,font=small,labelfont=bf}
\caption{ The average and maximum percentages off the optimal for all configurations with certain number of groups. In these experiments, $LBA / PBA = 0.7$.  }
 \label{fig:mix_groups}
\end{figure}

\begin{figure}
\includegraphics[scale=0.25]{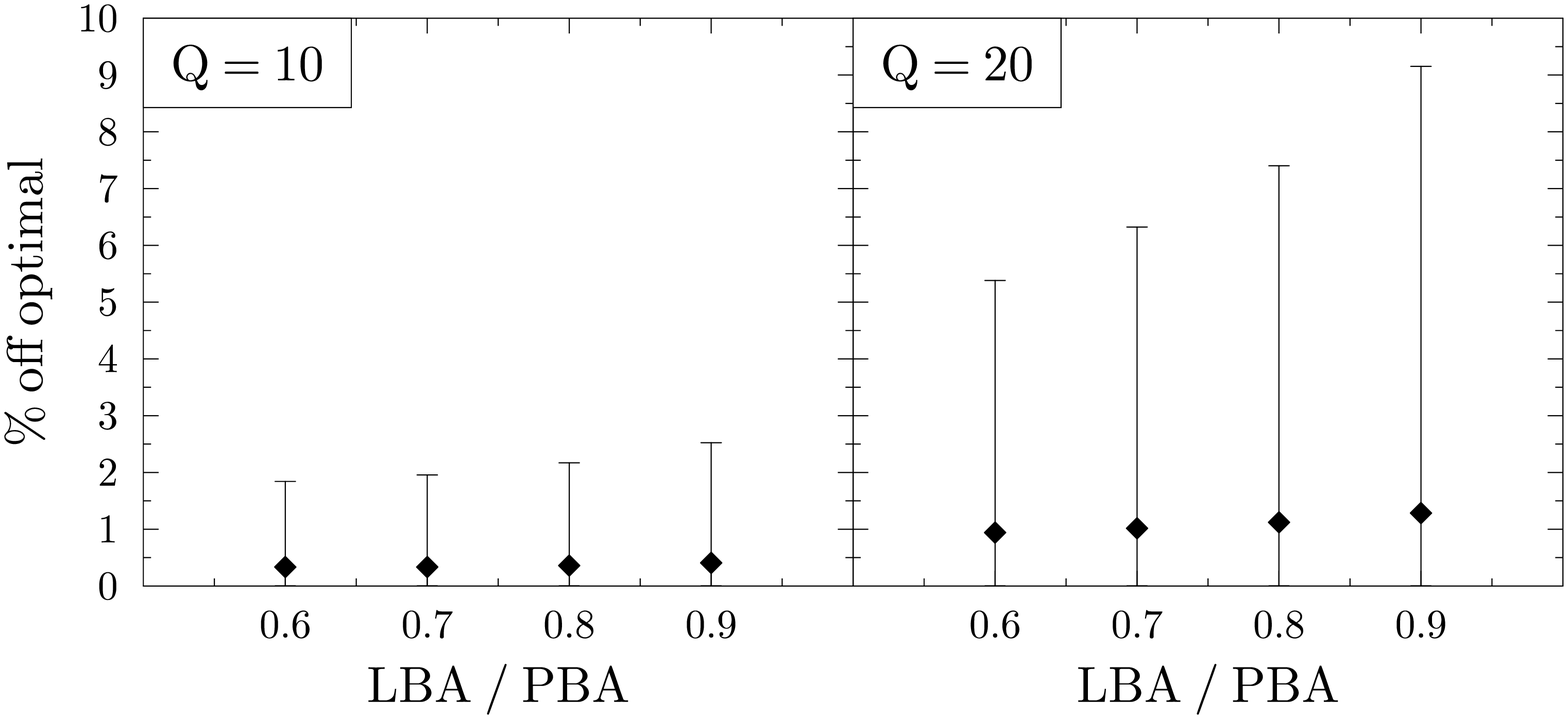}
\captionsetup{margin=10pt,font=small,labelfont=bf}
\caption{ The average and maximum percentages off the optimal for all configurations with certain values of LBA / PBA. The number of groups is fixed to 5. }
 \label{fig:mix_over}
\end{figure}

Trends from this analysis are displayed in Figures~\ref{fig:mix_groups} and \ref{fig:mix_over}. Both graphs show the average and maximum percentage differences of our policy from the optimum for the two different values of Q for every workload configuration tried. On average, our policy is below 1\% of the optimal for all the workload configurations. The maximum departures from the optimal are between 2\% (when Q=10) and 9\% (when Q=20). Put differently, the closed-form we propose works well unless for very skewed workloads where a group is either very hot or very cold. 

In order to take such skew into account in practice, we provide special treatment to the coldest group. When the hit rate of the coldest group is less than a certain fraction of the second coldest group, set to $5\%$ in our implementation, we allocate a fixed over-provisioned space to the coldest group, in our implementation $5\%$ of the smallest group's logical size. For the remaining groups we allocate over-provisioned space using the closed-form method. This works in practice well for the TPC-C Workload, whereon a high percentage of the data is extremely cold relative to the rest. 
 
\subsection{Temperature Detection} \label{sec:temp_det}

A key component of subroutine \ref{lst:writes} used to handle writes is to determine if the logical page should remain on the same group, or be demoted or promoted. A page should be promoted or demoted only if its update frequency changes relative to the group. Wolf uses one independent temperature detector for each group. If the temperature detector deems a page in group i relatively hotter or colder than its group, it migrates it to groups i+1 or i-1 respectively. 

Temperature detectors are subject to a trade-off between bookkeeping overhead and accuracy, and implementing a detector that strikes a good balance between these aspects is non-trivial. In our implementation, we used a detector inspired from \cite{bloom}. It consists of 2 bloom filters, one active and one passive. When a page is written to a group, its logical address is inserted into the active filter. Each group has independent write intervals, each of which is set to be as long as the number of pages in the group. At the end of an interval, the passive bloom filter is erased, the active one becomes passive, and a new active filter is created. This new bloom filter is initialized with 0.3 as the false probability rate and the projected number of elements is set to the current size of the group. Thus, each page in the SSD needs 5 bits in RAM for both bloom filters, which is not a significant overhead. The filters are used as follows. If the address of a page update exists in both active and passive filters and is an application update, the page is promoted to the next hotter group. If the address is in neither bloom filters and the write is a garbage-collection migration, the page is demoted to the next colder group. Otherwise, the page is updated in its current group. Designing a better temperature detector for Wolf is a topic for future work. 

\subsection{Garbage-Collection and Parallelism} \label{sec:Garbage-Collection-and-Parallelism}

Wolf exploits the SSD's available parallelism and achieves load-balancing by partitioning each group equally among all LUNs. It does so by maintaining free space for every group in every LUN, similarly to~\cite{adaptive}. We introduce the notion of a \textbf{subgroup} to refer to the blocks of given group that are on a particular LUN. Wolf's garbage-collection policy issues a garbage-collection operation in a subgroup as soon as the number of free pages in the subgroup falls below $B$. The manner in which a victim is chosen is described in section \ref{sec:cleaning_within_groups}. In terms of scheduling, a write targeting group $g_x$ is scheduled on some subgroup of $g_x$ that has free space and whose LUN is currently non-busy. During garbage-collection, the pages that are migrated can be written on any subgroup of the target group. 

\subsection{Wear-Levelling} \label{sec:wear-leveling}

The specifics of wear-levelling are orthogonal to the design of Wolf. Dynamic wear-levelling can take place by reassigning old blocks from hot groups to young groups to decelerate their wear, and young blocks from cold groups to hot groups to accelerate their wear. Static wear-levelling can take place in the coldest group to free exceptionally young blocks and assign them to hot groups.

\section{Evaluation}

Let us evaluate Wolf against the sate-of-the-art. As \cite{stoica}, we rely on simulation for our evaluation. We thus ignore the impact of error correction and bad block management on write-amplification, and the possible RAM limitations of a given SSD model. Rather, we focus on the impact of over-provisioned space utilisation on write-amplification with a simulator that reflects the internal structure of an actual SSD.  We are in contact with a SSD constructor to proceed to a validation of our simulation results. 

The recent scheme in FDP~\cite{stoica} has been shown to dominate existing block managers that strive to separate pages based on their update frequencies to reduce write-amplification, so it will serve as the main frame for comparison. We implemented in EagleTree the data allocation mechanisms from FDP, as described in~\cite{stoica}. We begin the evaluation by assuming an oracle temperature detector that knows the precise update probability of every page. We start by comparing how these schemes adapt to changes in workload. 



\subsection{Frequency Swap} 
\label{sub:changing_update_frequencies}

In the first experiment, we examine how the schemes adapt to a change in the update frequencies of two groups. Groups 1 and 2 are equally-sized and have update frequencies $10\%$ and $90\%$, and at some point their update frequencies swap. 

\begin{figure}
\includegraphics[scale=0.25]{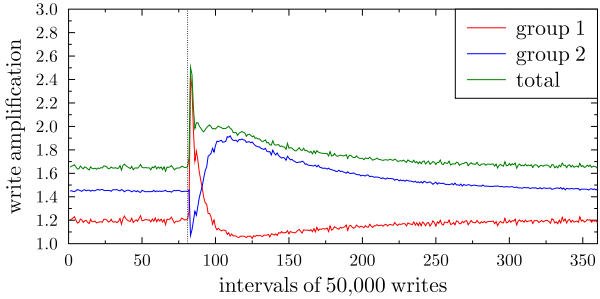}
\captionsetup{margin=10pt,font=small,labelfont=bf}
\caption{ How write-amplification evolves in time with FDP as we swap the update frequencies of two groups. The swap occurs at the dotted line. }
\label{fig:mass_migrations_prob}
\end{figure}

Figure \ref{fig:mass_migrations_prob} shows the evolution of write-amplification as a result of a swap with FDP. As described in Section \ref{sec:wolf}, the pages originally in group 2 which turned cold reside on blocks that occupy a significant amount of over-provisioned space, and it takes a considerable amount of time before the garbage-collection algorithm targets them . Thus, write-amplification is high and slow to decrease again. 

\begin{figure} 
\includegraphics[scale=0.25]{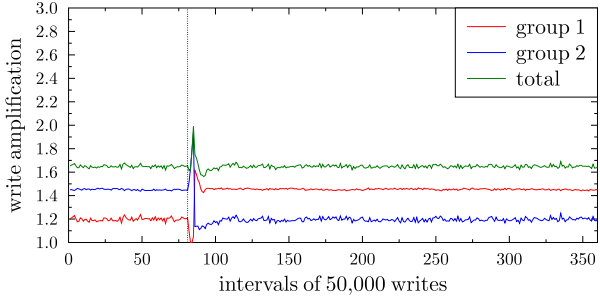}
\captionsetup{margin=10pt,font=small,labelfont=bf}
\caption{ How write-amplification evolves in time with Wolf as we swap the update frequencies of two groups. The swap occurs at the dotted line. }
\label{fig:mass_migrations_sol}
\end{figure}

In Figure~\ref{fig:mass_migrations_sol}, we see the result of a swap in Wolf. The change in update frequency is quickly detected, and movement operations aggressively take place in group 1 to reallocate over-provisioned space from the now cold group to the now hot group. Equilibrium is restored much more quickly.

We can compare the schemes using a baseline whereby no swap happens. The additional number of migrations for Wolf and FDP respectively relative to the no-swap scenario and divided by PBA is $0.7\%$ and $152.1\%$. In other words, Wolf hardly requires any additional migrations except a short spike, whereas FDP entails an additional total overwriting the SSD 1.5 times over. 

In order to generalize this result, we show the difference in performance between FDP and Wolf when the groups being swapped have different update frequencies. In the experiment, there are 5 groups with ids 0,1,2,3 and 4 that have exponentially increasing update frequencies of around $3.2\%$, $6.4\%$, $12.8\%$, $25.6\%$ and $51.2\%$, such that the sum of the update frequencies adds up to $100\%$. We swap the update frequencies of every pair of groups. We measure the total difference in migrations between FDP and Wolf and normalize it through division by the number of physical pages in the SSD. The bar chart shows that Wolf always outperforms FDP, with up to 2.2x performance improvements. The performance improvement increases with the difference between the groups in terms of update frequency. 

\begin{figure} 
\includegraphics[scale=0.6]{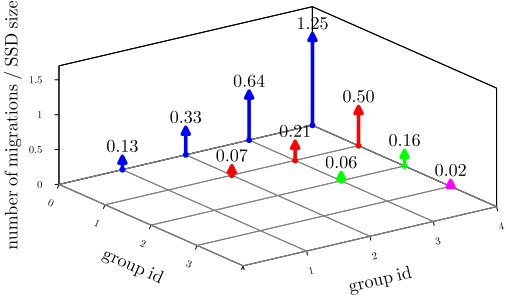}
\caption{ Garbage-collection performance as we swap the update frequencies of two groups after 5 million writes. }
\label{fig:swapping}
\end{figure}

\subsection{Realistic Workload} \label{sec:realistic}

In this section, we examine how Wolf behaves under a TPC-C workload. We generated an IO trace of the TPC-C benchmark using Shore-MT and Shore-Kits \cite{kits}. We added print statements into the code in Shore-MT to collect the logical addresses being written to. We used a scaling factor of 48 to match the size of our simulated SSD. The amount of RAM given to the Shote-MT buffer poll was set to  5$\%$ of the size of the data. 
 


A challenge of using TPC-C for benchmarking over SSDs is that the size of the database grows. Of every 100 writes on average, 3 are to new addresses. This is a problem as it causes over-provisioning to change throughout an experiment thereby making it difficult to control and interpret. 

\begin{figure} 
\includegraphics[scale=0.25]{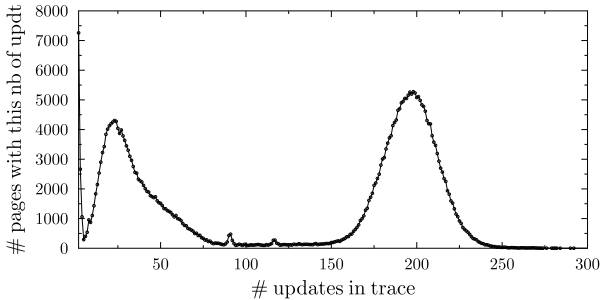}
\captionsetup{margin=10pt,font=small,labelfont=bf}
\caption{Update histogram of logical addresses in $\textup{TPC-C}_{\textup{init}}$ that have update frequencies between 1 and 300, which includes $46\%$ of the pages. Of the remaining pages, $54\%$ are never updated and $0.2\%$ are updated more than 300 times. }
\label{fig:skew}
\end{figure}

What is even more interesting is that the data that is added to the database throughout an experiment, which we coin $\textup{TPC-C}_{\textup{added}}$, has distinctly different workload characteristics than the data written during initialization, which we coin $\textup{TPC-C}_{\textup{init}}$. Figure \ref{fig:skew} shows the update frequency distribution for the logical pages in $\textup{TPC-C}_{\textup{init}}$. The temperature of most pages in $\textup{TPC-C}_{\textup{init}}$ remains stable. Of the pages in $\textup{TPC-C}_{\textup{added}}$, however, $68\%$ are only written once and never updated and the remaining $32\%$ tend to be hot for a short period of time and are then never updated again. Investigating how to cater for these different workload characteristics is the subject of future work. 

Here, we just focus on the updates in $\textup{TPC-C}_{\textup{init}}$ (47 million writes to $802816$ addresses). We assume that the updates in $\textup{TPC-C}_{\textup{added}}$ (10 million writes to $1815483$ addresses) are written elsewhere and so we remove them from the workload. Thus, over-provisioning remains constant throughout our experiments, and the workload is stable: the temperature of pages does not significantly change. 

How does Wolf compare to FDP under a TPC-C workload? When investigating this question, we found two factors that play a role: (i) the definition of groups, as their size and update frequency changes (adaptive for Wolf vs. non-adaptive for FDP), and (ii) the allocation of over-provisioned space (closed form for Wolf vs. the iterative method from \cite{Desnoyers, stoica}, which is theoretically optimal if group sizes and update frequencies are defined). In order to separate the influence of each factor, we ran two sets of experiments: first we varied group definition (both for closed form allocation), then we varied allocation strategy (both for adaptive groups).

\begin{figure} 
\includegraphics[scale=0.25]{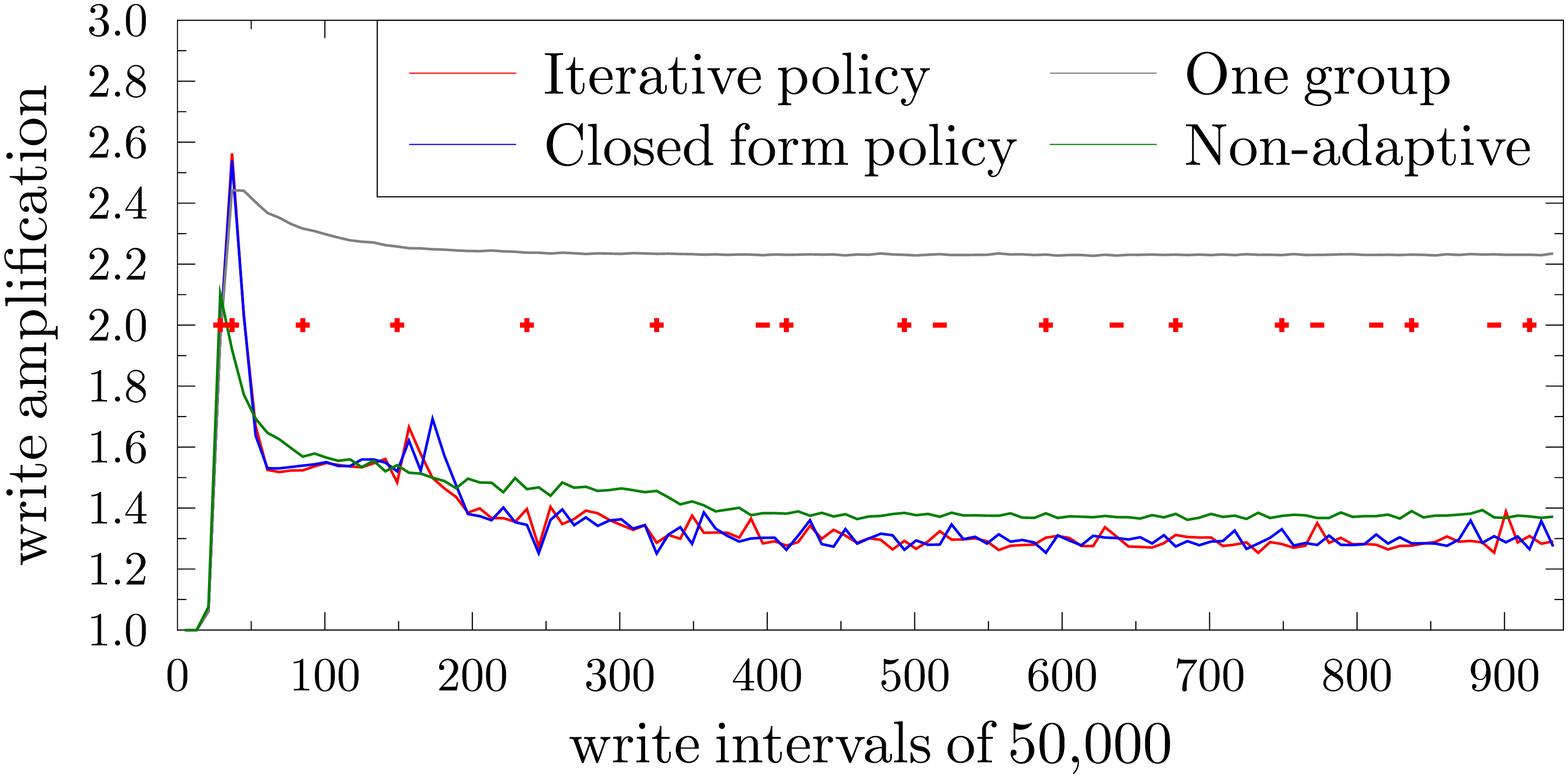}
\captionsetup{margin=10pt,font=small,labelfont=bf}
\caption{Comparison of two methods for allocating over-provisioned space to different groups. The red pluses and minuses indicate the creation or merging of groups respectively for the closed-form policy. }
\label{fig:iter-vs-closed}
\end{figure}

Figure \ref{fig:iter-vs-closed} shows the results of both experiments. The blue line corresponds to Wolf (flexible group definition, closed form allocation), the red line corresponds to the theoretically optimal iterative method (with flexible group definition), and the green line corresponds to FDP's group definition (with closed form allocation). The red and green line allow us to separate the features of FDP and how they relate to Wolf. The grey line, used as a baseline, illustrates write-amplification over time if all pages are mixed in one group. 

First, our experiment confirms the significant superiority of data allocation schemes that separate pages based on update frequency over the baseline that considers over-provisioned space as a single group.

Second, we observe that the policies corresponding to the red and blue lines perform similarly. The iterative policy involves $0.2\%$ less migrations than the closed-form policy, which suggests that for a realistic workload the theoretical difference between the policies explored in section \ref{sec:mixing} is in fact small. The number of groups in Wolf oscillates between 7 and 9 groups when running TPC-C. The reason for this is the skew of the workload. The last group tends to be so much hotter than the second hottest group, that Wolf continually creates additional hotter groups (according to the policy in \ref{sec:Dynamic_Group_Management}). However, this leads to the middle groups eventually converging in terms of hit rate, and they are thus merged. The creation and merging of groups does not involve any significant penalty. We observed some fluctuations arising in garbage-collection which occur due to surges in movement operations (but smoothed in Figure \ref{fig:iter-vs-closed}  for readability). Flattening such surges is the subject in future work. The mechanism we proposed in Section~\ref{sec:mixing} to handle coldness skew kicks in with the TPC-C workload, as the difference in update frequency between the coldest and second coldest groups is two orders of magnitude greater than the difference in update temperature between any other two adjacent groups. 

Third, we observe that the red and blue lines perform consistently better than the green line (by a approximately $22\%$ in equilibrium). For the red and blue lines, the update frequency of groups is continuously measured and adapted as described in section \ref{sec:Dynamic_Group_Management}. For the green line, we consider that groups have a fixed order, and that the pages in group i are twice as hot as pages in group i-1 and half as hot as pages in group i+1, as described in~\cite{stoica}. 
The workload skew explains the difference between the flexible group management from Wolf, and FDP's group management. The problem comes from FDP's assumption that the hit rate of one group is within a given factor of its adjacent groups. Wolf overcomes this by measuring how many writes actually target each group in each interval. These much more accurate figures of update frequencies allow the over-provisioning allocation mechanism to give different groups an allocation of over-provisioned space that yields slightly lower write-amplification. This shows that Wolf performs slightly better than the state-of-the-art for a realistic stable workload, while it outperforms it significantly when the workload changes. 

As far as we are aware, real SSD manufactures today don't implement schemes like FDP or Wolf. On real SSDs, pages with different update frequencies are mixed on the same blocks. Thus, running the experiments in this subsection on a real SSD would result in a much higher overall write-amplification that would resemble the gray line in figure \ref{fig:iter-vs-closed}. We hope that with the recent advent of schemes like FDP and Wolf, more SSD manufactures would indeed begin implementing such schemes.

\section{Conclusion} 
\label{sec:conclusion}

SSD write-amplification is a key aspect of their performance, which depends on (i) how much over-provisioned space is
made available, and (ii) how this over-provisioned space is allocated to different virtual partitions of the SSDs (groups) 
defined by their update frequency. 
In this paper, we presented Wolf, a FTL block manager, designed to adapt well to abrupt changes in workload and to the skew of realistic workloads such as TPC-C. Future work includes the deployment of Wolf in an open-channel SSD and its evaluation with a larger set of database workloads.


\bibliographystyle{abbrv}
\bibliography{bib}  

\begin{thebibliography}{10}

\bibitem{closed}
R.~Agarwal and M.~Marrow.
\newblock A closed-form expression for write amplification in nand flash.
\newblock In {\em GLOBECOM Workshops (GC Wkshps)}, pages 1846 -- 1850, 2010.

\bibitem{kits}
A.~Ailamaki, R.~Johnson, I.~Pandis, and P.~T\"{o}z\"{u}n.
\newblock Toward scalable transaction processing: Evolution of shore-mt.
\newblock {\em Proc. VLDB Endow.}, pages 1192--1193, Aug. 2013.

\bibitem{spikes}
P.~Bodik et~al.
\newblock Characterizing, modeling, and generating workload spikes for stateful
  services.
\newblock SoCC, pages 241--252, 2010.

\bibitem{uflip}
L.~Bouganim, B.~Jonsson, and P.~Bonnet.
\newblock {\em uFlip: Understanding Flash IO Patterns}, pages 1--12.
\newblock CIDR. 2009.

\bibitem{bux}
W.~Bux and I.~Iliadis.
\newblock Performance of greedy garbage collection in flash-based solid-state
  drives.
\newblock {\em Perform. Eval.}, pages 1172--1186, Nov. 2010.

\bibitem{adaptive}
L.-P. Chang and T.-W. Kuo.
\newblock An adaptive striping architecture for flash memory storage systems of
  embedded systems.
\newblock RTAS, pages 187--, 2002.

\bibitem{dac}
M.-L. Chiang, P.~C.~H. Lee, and R.-C. Chang.
\newblock Using data clustering to improve cleaning performance for plash
  memory.
\newblock {\em Softw. Pract. Exper.}, pages 267--290, Mar. 1999.

\bibitem{EagleTree}
N.~Dayan, M.~K. Svendsen, M.~Bj{\o}rling, P.~Bonnet, and L.~Bouganim.
\newblock Eagletree: Exploring the design space of ssd-based algorithms.
\newblock {\em Proc. VLDB Endow.}, pages 1290--1293, Aug. 2013.

\bibitem{Desnoyers}
P.~Desnoyers.
\newblock Analytic modeling of ssd write performance.
\newblock In {\em SYSTOR}, pages 12:1--12:10, 2012.

\bibitem{dftl}
A.~Gupta, Y.~Kim, and B.~Urgaonkar.
\newblock Dftl: A flash translation layer employing demand-based selective
  caching of page-level address mappings.
\newblock In {\em ASPLOS}, pages 229--240, 2009.

\bibitem{limit}
R.~Haas and X.~Hu.
\newblock The fundamental limit of flash random write performance:
  Understanding, analysis and performance modelling.
\newblock Technical report, IBM, 2010.

\bibitem{write-amp}
X.-Y. Hu, E.~Eleftheriou, R.~Haas, I.~Iliadis, and R.~Pletka.
\newblock Write amplification analysis in flash-based solid state drives.
\newblock In {\em SYSTOR}, pages 10:1--10:9, 2009.

\bibitem{container}
X.-Y. Hu, R.~Haas, and E.~Evangelos.
\newblock Container marking: Combining data placement, garbage collection and
  wear levelling for flash.
\newblock In {\em Proceedings of the 2011 IEEE 19th Annual International
  Symposium on Modelling, Analysis, and Simulation of Computer and
  Telecommunication Systems}, MASCOTS '11, pages 237--247, 2011.

\bibitem{fast}
S.-W. Lee, W.-K. Choi, and D.-J. Park.
\newblock Fast: An efficient flash translation layer for flash memory.
\newblock In {\em EUC Workshops}, pages 879--887, 2006.

\bibitem{Lee2013}
Y.-S. Lee, S.-H. Kim, J.-S. Kim, J.~Lee, C.~Park, and S.~Maeng.
\newblock {OSSD: A case for object-based solid state drives}.
\newblock {\em 2013 IEEE 29th Symposium on Mass Storage Systems and
  Technologies (MSST)}, pages 1--13, May 2013.

\bibitem{improved}
X.~Luojie and B.~M. Kurkoski.
\newblock An improved analytic expression for write amplification in nand
  flash.
\newblock ICNC, pages 497 -- 501, 2012.

\bibitem{Ouyang2011}
X.~Ouyang, D.~Nellans, R.~Wipfel, D.~Flynn, and D.~Panda.
\newblock {Beyond block I/O: Rethinking traditional storage primitives}.
\newblock In {\em High Performance Computer Architecture (HPCA), 2011 IEEE 17th
  International Symposium on}, pages 301--311. IEEE, 2011.

\bibitem{bloom}
D.~Park and D.~H. Du.
\newblock Hot data identification for flash-based storage systems using
  multiple bloom filters.
\newblock MSST, pages 1--11, 2011.

\bibitem{log-fs}
M.~Rosenblum and J.~K. Ousterhout.
\newblock The design and implementation of a log-structured file system.
\newblock {\em ACM Trans. Comput. Syst.}, pages 26--52, 1992.

\bibitem{stoica}
R.~Stoica and A.~Ailamaki.
\newblock Improving flash write performance by using update frequency.
\newblock {\em Proc. VLDB Endow.}, pages 733--744, July 2013.

\bibitem{envy}
M.~Wu and W.~Zwaenepoel.
\newblock envy: A non-volatile, main memory storage system.
\newblock In {\em ASPLOS VI}, pages 86--97, 1994.

\end{thebibliography}

\appendix
\section{Deeper analysis} \label{App:AppendixA}

Through a series of simple manipulations and application of the Lambert W function, we can express equation \ref{eq:eq_6} in terms of $\delta$: 


\begin{equation} \label{eq:new_eq_6}
\begin{split}
 \frac{LBA}{PBA} & = \frac{\delta - 1}{\ln(\delta)}  \\ - \frac{PBA}{LBA} & = \ln(\delta) - \frac{PBA}{LBA} \delta  \\
e^{-PBA/LBA} & = e^{\ln(\delta) - \delta(PBA/LBA)} \\
e^{-PBA/LBA} & = \delta e^{-\delta(PBA/LBA)} \\
-\frac{PBA}{LBA} e^{-PBA/LBA} & = -\delta \frac{PBA}{LBA}  e^{-\delta(PBA/LBA)} \\
-\delta \frac{PBA}{LBA} & = W \left( -\frac{PBA}{LBA} e^{-PBA/LBA} \right) \\
\delta & = - \frac{LBA}{PBA} \cdot W \left( -\frac{PBA}{LBA} e^{-PBA/LBA} \right)
\end{split}
\end{equation} 

This is in fact the equation derived through different means in \cite{Desnoyers, stoica, improved}, so these analyses are equivalent.

\end{document}